\documentclass[12pt]{article}
\usepackage{graphicx}

\def\lsim{\mathrel{\raise.3ex\hbox{$<$\kern-.75em\lower1ex\hbox{$\sim$}}}} 
\def\gsim{\mathrel{\raise.3ex\hbox{$>$\kern-.75em\lower1ex\hbox{$\sim$}}}}

\begin{document} 

\renewcommand{\thefootnote}{\fnsymbol{footnote}}
\renewcommand{\arraystretch}{0.5}

%@twocolumnfalse\endcsname 
 
\mbox{ } \\[-1cm]
\mbox{ }\hfill TUM--HEP--469-02\\%[-1mm]
\mbox{ }\hfill MADPH--02--1283\\%[-1mm]
\mbox{ }\hfill hep--ph/0207133 v.2\\%[-1mm]
\mbox{ }\hfill \today\\%[-1mm]

\begin{center}
  {\large\bf SUSY In The Sky: Observing Ultrahigh Energy Cosmic Neutralinos}\\[6mm]
  Cyrille Barbot$^1$, Manuel Drees$^1$, Francis Halzen$^2$ and
 Dan Hooper$^2$ \\[6mm] 
  {\it $^1$Physik Dept., TU M\"unchen, James Franck Str., D-85748
    Garching, Germany\\[6mm]
    $^2$Department of Physics, University of Wisconsin, 1150
    University Avenue, Madison, WI~53706, USA}
\end{center} 

%\maketitle 
 
\begin{abstract}
In models where the ultra-high energy cosmic ray problem is solved by
top-down scenarios, a significant flux of ultra-high energy
neutralinos is predicted.  We calculate the number of events expected
from such particles in future experiments such as EUSO or OWL.  We
show that by using the Earth as a filter, showers generated by
neutralinos can be separated from neutrino generated showers.  We find
that for many models, observable rates are expected.
\end{abstract} 
%\pacs{11.30.Pb, 14.80.Ly, 95.85.Ry, 96.40.De, 96.40.Pq, 96.40.Tv} 

%\maketitle
%\thispagestyle{empty}

\section{Introduction}

\setcounter{footnote}{1}

Cosmic ray observations have determined that the spectrum of the
highest energy cosmic rays extends beyond $10^{20}\,$ eV
\cite{hiresagasa}.  Observations have also indicated that the highest
energy spectrum is dominated by protons rather than photons
\cite{protons}. Above $\sim 5 \times 10^{19}\,$ eV, protons can
interact with cosmic background photons at the $\Delta$-resonance
generating pions.  Above this energy, called the GZK
(Greisen-Zatsepin-Kuzmin) cutoff \cite{gzk}, the proton energy loss
length is near 50 Mpc, thus requiring semi-local sources to produce
the observed flux.  The lack of any such known sources has spawned a
great deal of speculation as to the origin of these particles.  A
common class of models, called top-down scenarios, involve
supermassive particles which decay or annihilate generating the
highest energy cosmic rays \cite{top,Sigl}.

The decay of superheavy particles has been studied in some detail
\cite{bere1,XDM,frag,drees,toldra}. In particular, it has been
demonstrated that a significant amount of the initial energy of
such a particle can be emitted in the form of ultra-high energy
supersymmetric particles \cite{bere1,drees,toldra}. In most models,
the lightest supersymmetric particle is a neutralino. This neutralino,
weakly interacting and stable by virtue of $R-$parity, can travel
cosmological distances without absorption or scattering. In this
paper, we discuss the prospects for observing ultra-high energy cosmic
neutralinos in future very large area, satellite-borne air shower
experiments.

\section{Ultra-High Energy Fragmentation To Neutralinos}

In the general framework of top-down scenarios, one has to consider
the decay of super-heavy $X$ particles with a mass of the order of
$10^{21}$ to $10^{25}$ eV, and a lifetime comparable to or longer than
the age of the universe. Such a long lifetime can be ensured by
``storing'' the $X-$particles in cosmological defects, which can
survive into the present epoch \cite{top,defects}. Alternatively, free
$X-$particles might be long--lived since their decay is suppressed,
e.g. by (approximate) symmetries \cite{freeX}. For a review of
different candidates, see \cite{Sigl}. Such particles could be
produced in the very early times of the universe, e.g. at the end of
inflation \cite{produce}. The typical decay modes of the $X$ particles
are generally unknown and/or quite model dependent. However, if there
is no additional ``new physics'' scale between $M_{\rm SUSY} \sim 1$
TeV and $M_X$\footnote{This hypothesis, known as the ``desert
hypothesis'', is well motivated by the fact that the existence of new
physics between the GUT scale and the SUSY breaking scale would
destroy the very impressive feature of ``natural'' unification of the
gauge couplings at $M_{GUT} \sim 10^{16}$ GeV occuring in the MSSM
\cite{gut}.}, the $X$ particles should decay into $N$ ``known''
particles of the minimal supersymmetric standard model (MSSM), and
usual particle physics allows us to study in detail the shower
generated by the primary products of the initial $X$ decay.

A detailed computation of the spectra of stable particles (protons,
photons, neutralino LSPs, electrons and neutrinos of the three
species) obtained in such decay showers has been described in
\cite{drees,bd2}. We recall here that at the energies we are
considering, it is necessary to take into account all the gauge
couplings of the MSSM; indeed, at the scale of unification, they are
all of the same strength, so that electroweak (and some Yukawa)
interactions can be as relevant as the QCD ones. The perturbative part
of the shower was computed by solving numerically the complete set of
evolution equations \cite{bd2} for the relevant fragmentation
functions of the MSSM. We carefully modeled the decays of unstable
particles with mass near $M_{\rm SUSY} \sim 1$ TeV, as well as the
hadronization process at the GeV scale for light quarks and gluons. We
found that the LSP flux depends only mildly on the spectrum of
superparticles, as long as the LSP is a bino--like neutralino
\cite{bd2}. Some sample spectra are shown in Fig.~1.\footnote{The
primary 10--body decay $X \rightarrow 5q 5 \tilde q$ has been modeled
using phase space only, i.e. ignoring any possible dependence of the
matrix element on external momenta.}

Here we have conservatively assumed that $X$ particles have an
overdensity of $10^5$ in the vicinity of our galaxy, as expected
\cite{XDM} for $X$ particles that move freely under the influence of
gravity.\footnote{The exact profile of the halo of $X$ particles does
not affect our results as long as most UHECR events originate at
distances well below one GZK interaction length.} This minimizes the
expected neutralino flux, since all scenarios are normalized by
matching \cite{topdown} the predicted proton spectrum at $E \simeq
10^{20}$ eV to the highest energy cosmic ray observations. Data
\cite{protons} indicate that most UHE events have protons (or heavier
ions, which however cannot be produced in top--down models), not
photons, as primary particles. Specifically, Yakutsk, Haverah Park
and, most recently, AGASA observe more muons in the events than
expected from photon primaries. The longitudinal development of the
most energetic Fly's Eye event indicates that it is not due to a
photon. Finally, the highest energy AGASA events should show a
North--South asymmetry if they were due to photons (which, at that
energy, initiate an electromagnetic cascade already in the Earth's
magnetosphere), but no such asymmetry is observed. On the other hand,
top--down models predict the photon flux at source to exceed the
proton flux \cite{bd2}. 

Following ref.\cite{topdown} we are thus compelled to assume that most
UHE photons are absorbed somewhere between source and Earth. According
to ref.\cite{Sigl} this is in fact expected if most sources are at
cosmological distances. At $E \simeq 10^{20}$ eV the attenuation
length for photons is at least one order of magnitude shorter than
that for protons. This ratio is closer to 50, if the extragalactic
$\vec{B}$ fields average at least $10^{-9}$ G, as is widely
expected. On the other hand, according to standard estimates
\cite{Strong} the density of radio photons in our galaxy is only about
one order of magnitude higher than the average value in intergalactic
space, leading to an attenuation length of $10^{20}$ eV photons of at
least 100 kpc. In order to obtain near--complete absorption of such
photons in our galaxy the actual density of radio photons would have
to be at least ten times higher than the accepted value. This may not
be impossible. The galactic (and extragalactic) radio background is
reasonably well known only at frequencies above 100 MHz. However, we
are most interested in photons with frequency of only a few MHz, which
have the highest cross section for $e^+e^-$ pair production with
$10^{20}$ eV photons. This frequency band is difficult to observe on
Earth, due to strong absorption in the ionosphere.

\begin{figure}[h!]
\centering\leavevmode
\includegraphics[width=5in]{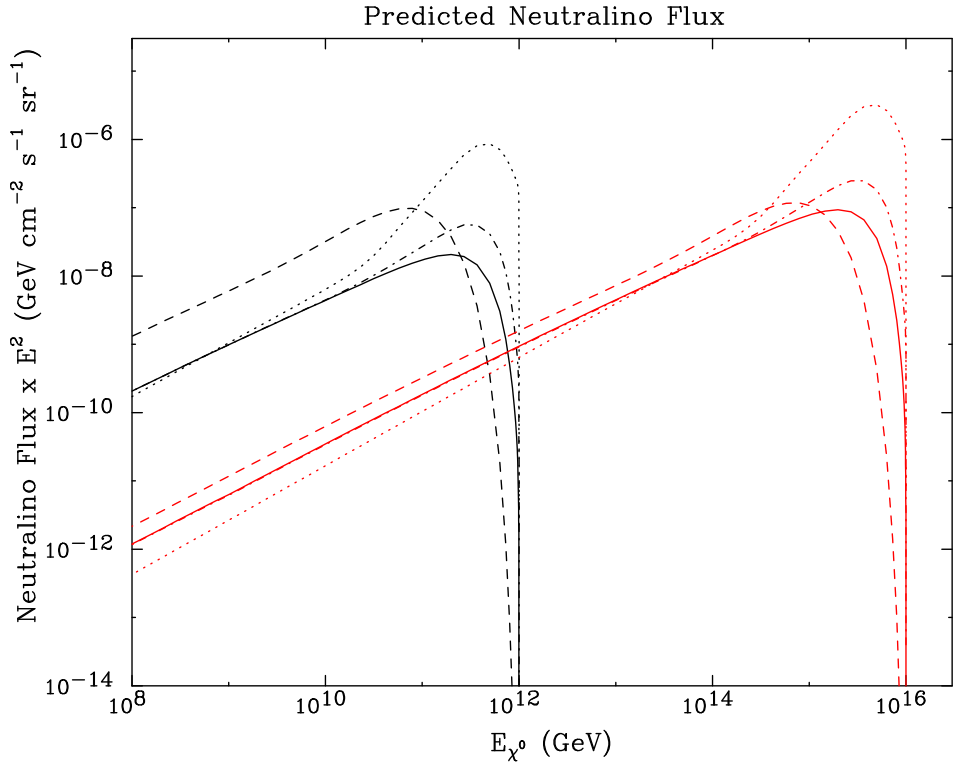}
\caption{The spectrum of neutralino LSP's predicted for the decay of
superheavy particles with mass $M_X = 2 \cdot 10^{21}$ eV (left set of
curves) and $M_X= 2 \cdot 10^{25}$ eV (right) normalized
\cite{topdown} by the proton spectrum to the ultra-high energy cosmic
ray flux, for a ``galactic'' distribution of sources where most UHECR
events originate from $X$ decays in the halo of our galaxy. For a
homogeneous distribution, the spectrum is enhanced by up to a factor
of 15. Spectra are shown for primary $X$ decays into quark$+$antiquark
(solid), quark$+$squark (dot-dash), $SU(2)$ doublet lepton$+$slepton
(dots) and 5 quark$+$5 squark (dashes).  Note that for the case of
$M_X = 2 \cdot 10^{21}$ eV decays, the spectrum peaks in the energy
range most accessible to air shower experiments.}
\label{two}
\end{figure}

\section{Signatures of Ultra-High Energy Neutralinos}

Ultra--relativistic neutralinos interact with quarks by $t-$channel
$Z$ and $W^{\pm}$ exchange, as well as by the exchange of squarks in
the $s-$ or $u-$channel. These interactions either directly yield an
LSP, or produce a heavier neutralino or chargino which quickly decays
to the lightest neutralino (except, perhaps, in the case of
near--degenerate masses). Either interaction generates a shower which
can be observed by air shower experiments.

The background for this signal consists of showers generated by
ultra-high energy cosmic neutrinos.  The neutrino interaction length
becomes comparable to the radius of the earth around $10^5$ GeV.  By
$10^9$ GeV, only about one out of 1000 neutrinos passes through the
Earth without interaction (see figure 2). A neutralino, however,
depending on the choice of SUSY parameters, will have a different
interaction cross section and, therefore, different absorption
properties. The size of this cross section depends sensitively on the
neutralino eigenstate, which in general is a composition of bino, wino
and neutral higgsinos. A wino-- or higgsino--like neutralino has
couplings to $W$ and/or $Z$ bosons that resemble or even exceed those
of neutrinos. In contrast, a bino--like neutralino has very small
couplings to gauge boson, because its superpartner, the $U(1)_Y$ gauge
boson, does not couple to other gauge bosons. The couplings of
bino--like neutralinos to squarks are of full $U(1)_Y$ gauge strength,
but squark searches at the Tevatron \cite{tevsearch} tell us that
first and second generation squarks must be at least three times
heavier than $W$ bosons. Note also that models with radiative breaking
of the electroweak gauge symmetry prefer the lightest neutralino to be
bino--like in most of parameter space \cite{binomodel}. Typical
parameter choices therefore predict neutralino-nucleon cross sections
one or two orders of magnitude smaller than neutrino-nucleon cross
sections \cite{bere1}. With a significantly smaller cross section,
very high energy cosmic neutralinos may travel through the Earth
producing upgoing events at much higher energies than
neutrinos. Upgoing showers with energy above 100 PeV or so would be a
smoking gun for cosmic neutralinos.

Furthermore, by virtue of $R-$parity, neutralinos will generate less
energetic neutralinos in each interaction, thus not depleting their
number.  Tau neutrinos also display this property \cite{saltzberg},
but not as dramatically.  The difference comes from the fact that high
energy tau leptons lose energy in propagation whereas charginos decay
quickly enough to lose very little energy in propagation. Also, phase
space arguments indicate that a larger fraction of a decaying
chargino's energy goes into the resulting (massive) neutralinos than a
decaying tau's energy goes into the new (essentially massless) tau
neutrino.  Together, these effects indicate that tau regeneration is
largely ineffective above about $10^8$ GeV.  On the other hand, for
even moderately smaller neutralino cross sections, the Earth can
remain effectively transparent to cosmic neutralinos at much higher
energies.

Our calculations of tau neutrino and neutralino regeneration in the
Earth were done with a Monte Carlo simulation which, at each
interaction, calculated the energy lost in the interaction and
following propagation \cite{saltzberg}. Our treatment of $\tau$
propagation includes $e^+e^-$ pair production, photonuclear
interactions, bremsstrahlung and ionization energy losses. As stated
earlier, any unstable superparticle produced in LSP interactions is
too short--lived to lose energy prior to its decay. We estimate that
each interaction, if necessary followed by superparticle decay, will
reduce the energy of the LSP by slightly more than a factor of two;
this effect is included in our treatment of LSP regeneration. Our code
demonstrated the appearance of a `pile-up' of outgoing particles at an
energy corresponding to an interaction length equal to the size of the
Earth.  For tau neutrinos, this occurs at PeV energies, but can be
considerably higher for neutralinos, due to their smaller cross
section.

%\vskip 0.5cm
 
\begin{figure}[h!] 
\centering\leavevmode
\includegraphics[width=4.5in]{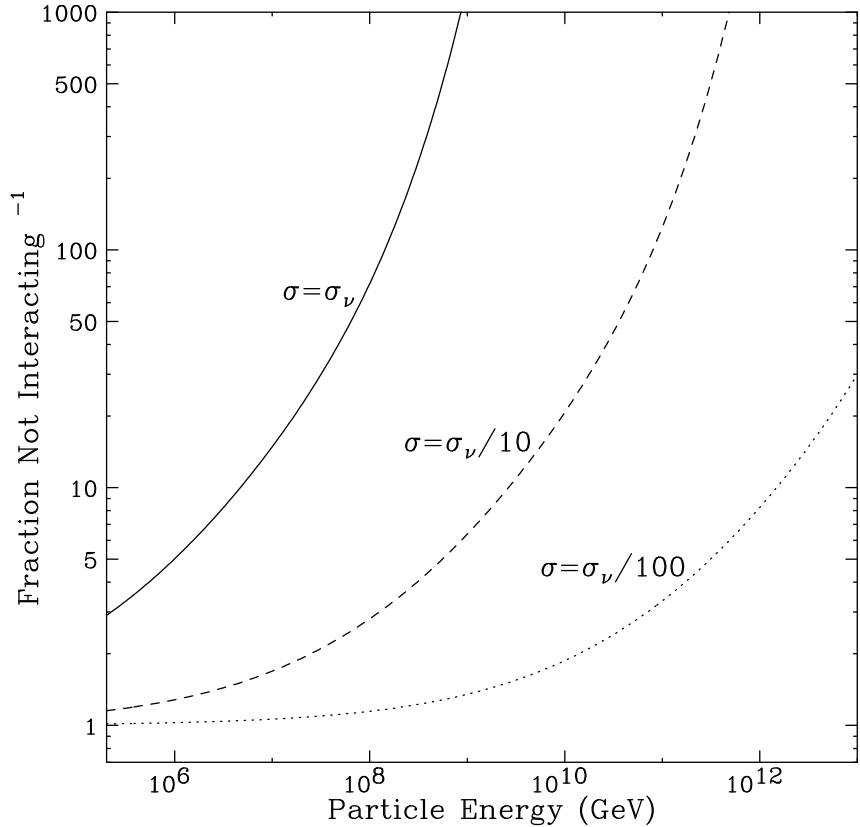}
\caption{The fraction of neutrinos or neutralinos which pass through
the Earth (integrated over zenith angle less than 85 degrees) as a
function of energy. Results are shown for particles with total cross
sections with nucleons equal to that for neutrinos as well as for
particles with cross sections ten and one hundred times smaller.
Regeneration effects are not included (see end of sec.~3).}
\label{one} 
\end{figure} 

\section{Prospects For Detection In Air Shower Experiments}

The flux of very high energy neutralinos from top-down scenarios can
be calculated assuming that this is the mechanism which generates the
highest energy cosmic rays \cite{top,frag,drees,bd2}. Given a
sufficient cosmic flux, these neutralinos may be detected in future
air shower experiments. The challenge, however, is not merely
observing the showers generated in neutralino interactions but in
differentiating these cosmic neutralinos from neutrinos.

We have calculated the number of neutralino events predicted for a
variety of top-down models associated with the highest energy cosmic
rays in a future experiment such as EUSO \cite{EUSO} or OWL
\cite{OWL}. EUSO and OWL are proposed satellite experiments which
observe fluorescence in the Earth's atmosphere generated in very high
energy showers.  Such experiments are expected to observe on the order
of 150,000 square kilometers of surface area on the Earth.  Particles
which pass through the Earth can interact in the shallow Earth or
atmosphere generating upgoing showers observable by fluorescence or
Cerenkov radiation.  Ultra-high energy showers reach a maximum near a
slant depth of 850 $\rm{g}/\rm{cm}^2$, corresponding to a depth of 8.5
meters in water. Including the effective slant depth of the lower
atmosphere extends this to $\sim 0.015 \, $ km, thus providing a water
equivalent effective volume of $\sim150,000 \times 0.015 \sim 2250 \,$
cubic kilometers, a truly enormous volume. Such an experiment will be
capable of measuring both the energy and the direction of an observed
particle. 

Estimating the rate of neutrino--induced ``background'' events is
difficult at present since the neutrino flux at $E \gsim 10^9$ GeV is
not known. The flux of atmospheric neutrinos is completely negligible
at these energies. However, most proposed explanations of the UHECR
events also predict a significant UHE neutrino flux. We therefore use
the neutrino flux predicted by top--down models \cite{topdown} to
estimate the neutrino background. Fig.~3 compares signal and
background at $E \geq 1$ EeV for one such model, where we assume a
galactic distribution of $X$ particles, with primary $X \rightarrow q
\bar q$ decay and $M_X = 2 \cdot 10^{12}$ GeV. We see that signal and
background clearly have very different angular distributions even for
the larger LSP--nucleon cross section of $\sigma_\nu/10$. Regeneration
effects are included, but they cannot produce neutrino events at large
energy {\em and} large angle. Requiring the events to emerge more than
$5^\circ$ below the horizon removes almost all the background, with
little loss of signal; in the case at hand, we expect about 2 signal
events per year, compared to 0.1 background event. If the LSP--nucleon
cross section is smaller, a somewhat stronger angular cut may be
advantageous; on the other hand, at even higher energies it might be
better to use a slightly weaker cut. However, this variation of the
angular cut has negligible effect on the predicted signal rate,
compared to the uncertainty inherent in our estimates. In the
following we therefore apply a fixed angular cut of $5^\circ$ on the
signal in all cases. This cut will have to be optimized once the
angular resolution of the experiment is known. Moreover, measurements
at neutrino telescopes as well as AUGER should soon greatly improve
our knowledge of the neutrino flux at very high energies. Finally,
this figure also shows that a measurement of the angular distribution
of the signal will allow to determine the LSP scattering cross
section: for the larger cross section shown, there will be very few
vertically upgoing events. The dependence of the angular distribution
of the signal on the cross section becomes even more pronounced at
higher energies.

%\vskip 0.5cm
 
\begin{figure}[h!] 
\centering\leavevmode
\includegraphics[width=4.5in]{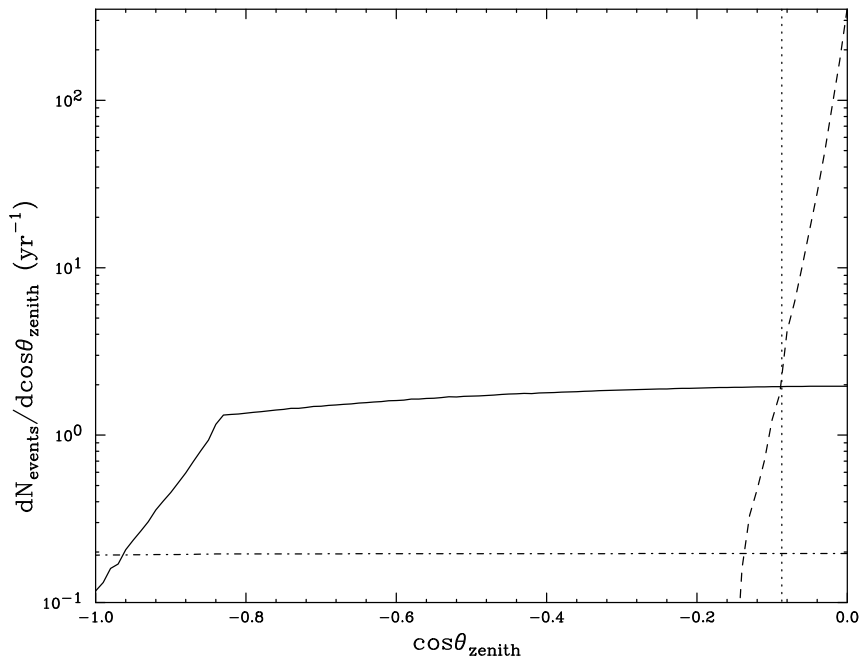}
\caption{The neutrino background (dashed) and LSP signal (solid:
$\sigma_{\rm LSP} = \sigma_\nu/10$; dot--dashed: $\sigma_{\rm LSP} =
\sigma_\nu /100$) at $E > 1$ EeV. Both signal and background result
from $X \rightarrow q \bar q$ decays of $2 \cdot 10^{12}$ GeV $X$
particles with a galactic distribution. The vertical dotted line
indicates the angular cut of $5^\circ$ applied to the signals listed
in table~1.}
\label{three} 
\end{figure}

\begin{table}[h!]
\begin{center}
\begin{tabular}{|c||c|c|}\hline \hline
$E_{\chi^0} \ge 1 \,\rm{EeV}$ & $\sigma_{\chi^0} =
\sigma_{\nu}/10$ & $\sigma_{\chi^0} = \sigma_{\nu}/100$  \\   
\hline \hline
$q \bar{q}$, $10^{21}$ eV, Galactic & 1.86 & 0.196 \\
$q \tilde{q}$, $10^{21}$ eV, Galactic & 2.96 & 0.306 \\
$5 \times q \tilde{q}$, $10^{21}$ eV, Galactic & 4.05 & 0.436\\
$l \tilde{l}$, $10^{21}$ eV, Galactic & 28.0 & 2.81\\
\hline  \hline 
$q \bar{q}$, $10^{25}$ eV, Galactic & 0.187 & 0.0189\\
$q \tilde{q}$, $10^{25}$ eV, Galactic & 0.213 & 0.0216\\
$5 \times q \tilde{q}$, $10^{25}$ eV, Galactic & 0.213 & 0.0216\\
$l \tilde{l}$, $10^{25}$ eV, Galactic & 0.615 & 0.0617\\
\hline  \hline 
$q \bar{q}$, $10^{21}$ eV, Homogeneous & 27.9 & 2.94 \\
$q \tilde{q}$, $10^{21}$ eV, Homogeneous & 44.4 & 4.56\\
$5 \times q \tilde{q}$, $10^{21}$ eV, Homogeneous & 60.8 & 6.54\\
$l \tilde{l}$, $10^{21}$ eV, Homogeneous & 420.0 & 42.15\\
\hline  \hline 
$q \bar{q}$, $10^{25}$ eV, Homogeneous & 2.81 & 0.284\\
$q \tilde{q}$, $10^{25}$ eV, Homogeneous & 3.20 & 0.324\\
$5 \times q \tilde{q}$, $10^{25}$ eV, Homogeneous & 3.20 & 0.324\\
$l \tilde{l}$, $10^{25}$ eV, Homogeneous & 9.23 & 0.926 \\
\hline \hline \hline
$E_{\chi^0} \ge 100\, \rm{EeV}$ & $\sigma_{\chi^0} =
\sigma_{\nu}/10$ & $\sigma_{\chi^0} = \sigma_{\nu}/100$ \\ 
\hline \hline
$q \bar{q}$, $10^{21}$ eV, Galactic & 0.0976 & 0.0344\\
$q \tilde{q}$, $10^{21}$ eV, Galactic & 0.391 & 0.122\\
$5 \times q \tilde{q}$, $10^{21}$ eV, Galactic & 0.0161 & 0.00716 \\
$l \tilde{l}$, $10^{21}$ eV, Galactic & 10.1 & 2.38\\
\hline  \hline
$q \bar{q}$, $10^{25}$ eV, Galactic & 0.0946 & 0.0143\\
$q \tilde{q}$, $10^{25}$ eV, Galactic & 0.116 & 0.0169\\
$5 \times q \tilde{q}$, $10^{25}$ eV, Galactic & 0.103 & 0.0159\\
$l \tilde{l}$, $10^{25}$ eV, Galactic & 0.435 & 0.0576\\
\hline  \hline
$q \bar{q}$, $10^{21}$ eV, Homogeneous & 1.46 & 0.516\\
$q \tilde{q}$, $10^{21}$ eV, Homogeneous & 5.87 & 1.83\\
$5 \times q \tilde{q}$, $10^{21}$ eV, Homogeneous & 0.242 & 0.107\\
$l \tilde{l}$, $10^{21}$ eV, Homogeneous & 151.5 & 35.7\\
\hline  \hline
$q \bar{q}$, $10^{25}$ eV, Homogeneous & 1.42 & 0.215 \\
$q \tilde{q}$, $10^{25}$ eV, Homogeneous & 1.74 & 0.254\\
$5 \times q \tilde{q}$, $10^{25}$ eV, Homogeneous & 1.55 & 0.239 \\
$l \tilde{l}$, $10^{25}$ eV, Homogeneous & 6.53 &0.864 \\
\hline \hline
\end{tabular}
\end{center}
\caption{Neutralino event rates per year in top-down scenarios in a
large area air shower experiment such as EUSO or OWL, with effective
volume $\simeq 2250$ cubic kilometers (water equivalent).  Rates are
shown for two choices of neutralino-nucleon cross sections, two
choices of energy threshold and several top-down models.  At the
energies considered, there is very little neutrino background for
upgoing events (see text).}
\label{table:I} 
\end{table}

Table 1 shows signal event rates for two choices of energy threshold,
$E_{\chi^0} \ge$ 1 EeV and 100 EeV. We also show results for the
stronger cut on energy in order to illustrate that at least in some
cases the LSP spectrum should be measurable over a significant range of
energies. The first case shown in the table corresponds to the
situation depicted in Fig.~3. Of course, the choice of a 100 EeV
threshold is even more effective in reducing the background, to the
level of $10^{-3}$ events per year. From the physics point of view an
energy threshold of 100 EeV should only be necessary in the unlikely
case that the total background of ultra--high energy neutrinos is
dominated by some mechanism not related to the observed UHECR events.
Regarding the energy threshold which can be achieved experimentally,
it has been argued that for upgoing events, the threshold could be as
small as a PeV \cite{thresh}.
 
The rates shown in table 1 are for a variety of primary $X$ decay
modes, and for ``galactic'' and homogeneous distributions of $X$
particles. It seems highly unlikely that $X$ particles will indeed be
distributed homogeneously, but it is conceivable that the majority of
sources contributing to the LSP flux is at cosmological distances
(e.g. if the $X$ particles are embedded in topological defects); the
homogeneous distribution is meant to be representative for such
models. Our results show that the $X$ distribution throughout the
universe has significant impact on the expected size of our signal.
For a full description of these models, see our previous paper
\cite{topdown}. We note that the neutralino signal is more sensitive
to the primary $X$ decay mode than the neutrino signal analyzed in
\cite{topdown} is. Not surprisingly, scenarios with (at least) one
superparticle in the primary decay produce a higher neutralino flux
than models where $X$ only decays into quarks.  Moreover, leptonic $X$
decays increase the predicted neutralino flux by another order of
magnitude, since in this case relatively few protons are produced,
leading to a higher source density required to explain the observed
UHECR events. On the other hand, choosing $M_X = 2 \cdot 10^{25}$ eV
rather than $2 \cdot 10^{21}$ eV significantly reduces the predicted
flux. Note, however, that in this case $X$ decays can only describe
the UHECR flux above $\sim 10^{20}$ eV \cite{topdown}; events at a few
times $10^{19}$ eV then have to be produced by an as yet unknown
source. 

As stated earlier, we normalize the LSP flux by assuming that (almost)
all UHE photons are absorbed between source and Earth, as indicated by
experiment \cite{protons}. Since absorption of UHE photons in our
galaxy is speculative, we comment on how the expected signal is
changed if this evidence is ignored, i.e. if the observe UHECR
spectrum is normalized to the sum of photon and proton fluxes. The
predicted LSP event rate for galactic models with $M_X = 2 \cdot
10^{25}$ eV would go down by about a factor of 4. If $M_X = 2 \cdot
10^{21}$ eV, the predicted event rate would go down by a factor of 2
to 3 for hadronic primary $X$ decays, and by about an order of
magnitude for purely leptonic primary $X$ decay.\footnote{The UHE
neutrino background from $X$ decay would be reduced by the same
factor.} Note that this "uncertainty" in the predicted event rate from
taking refs.\cite{protons} seriously or not is comparable to the
variation between different primary $X$ decay modes. Finally, we
remind the reader that the UHECR spectra measured by AGASA and HiRes
differ significantly in the post--GZK region, where we normalize our
fluxes, leading to a corresponding uncertainty in our predicted
signal.

\section{Conclusions}

The cosmic neutralino flux predicted in top--down scenarios could
possibly provide an interesting test of both supersymmetry and GUT
scale particle physics. To identify any showers generated in future
experiments as being generated by cosmic neutralinos, they will need
to occur at energies and from directions at which neutrinos would be
absorbed by the Earth. We have calculated the event rates for a
variety of such models for a large area air shower experiment such as
OWL or EUSO. We find that for many scenarios, the event rate is large
enough to be observable in principle. We should mention here that our
estimates of annual event rates assume 100\% duty cycle. This is
clearly not realistic for any experiment based on optical
observations. However, planning for the kind of space--based
experiment we envision is still in its early stage; a smaller duty
cycle might be compensated by a larger area and/or a longer period of
observation. 

We believe that searching for UHE LSPs is very important, since it is
the only measurement that can {\em qualitatively} distinguish between
``top--down'' and the more conventional ``bottom--up'' explanations
for the observed UHE events: in bottom--up models superparticles can
only be produced in the collision of accelerated protons, so the UHE
LSP flux will be a tiny fraction [typically ${\cal O}(10^{-6})$ or
less] of the UHE neutrino flux, much too small to be observed in any
currently conceivable experiment. In contrast, a sizable UHE LSP flux
is a generic prediction of top--down models. Moreover, the neutralino
event rate turns out to be a far more sensitive probe of details of
the model than the flux of neutrinos with energy exceeding $\sim 1$
PeV \cite{topdown}. We therefore find it encouraging that the
observation of UHE LSPs along the lines suggested in this paper, while
certainly not easy, should at least be possible.

\bigskip

{\large \it Acknowledgments}: We thank Andrew Strong for a useful
discussion. This work was supported in part by a DOE grant
No. DE-FG02-95ER40896 and in part by the Wisconsin Alumni Research
Foundation. The work of M.D. was partially supported by the SFB375 of
the Deutsche Forschungsgemeinschaft.

\end{document}